\documentclass[reprint,aps,prd,onecolumn,notitlepage,showpacs,nofootinbib,preprintnumbers,superscriptaddress]{revtex4-1}
\usepackage{amsmath}
\usepackage{amsfonts}
\usepackage{amssymb}
\usepackage{latexsym}
\usepackage{color,xcolor}
\usepackage{graphicx}
\usepackage{bm}
\usepackage{epsfig}
\usepackage[english]{babel}
\usepackage{hyperref}
\usepackage{times}
\usepackage{comment}

\def\bk{\textrm{bulk}}
\def\bd{\textrm{boundary}}
\def\cn{\textrm{corner}}
\def\const{\textrm{const}}
\def\thecs{\theta_{CS}}

\begin{document}

\title{Action growth rates of black holes in the Chern-Simons modified gravity}
\author{Yu-Chen Ding}
\author{Towe Wang}
\email[Electronic address: ]{twang@phy.ecnu.edu.cn}
\affiliation{Department of Physics, East China Normal University,\\
Shanghai 200241, China\\ \vspace{0.2cm}}
\date{\today\\ \vspace{1cm}}
\begin{abstract}
Among many modified gravity theories, the Chern-Simons modified gravity stands out as one of the few examples whose Dirichlet boundary problem has been well studied. Known solutions to this theory include the Schwarzschild black hole and a slowly rotating black hole. Making use of the Dirichlet boundary term of this theory, we calculate the late-time action growth rates of the Wheeler-DeWitt patch for the Schwarzschild and the slowly rotating solutions. In light of the conjecture of complexity/action duality, the result has implications on the quantum complexity of a state in the dual field theory.
\end{abstract}


\maketitle



\section{Introduction}\label{sect-intro}
Ever since the end of the last century, the AdS/CFT duality has revealed more and more elegant connections between the boundary conformal field theory (CFT) and the gravity theory in the bulk asymptotic to the anti-de Sitter (AdS) spacetime. Recently, Ryu and Takayanagi proved that entanglement entropy of the quantum information on the boundary is equivalent to the area of the minimal codimension-two hypersurface that shares the same boundary of CFT interval \cite{Ryu:2006bv}. The discovery showed the area, which is a bulk quantity, is deeply connected to the entanglement entropy, namely a CFT quantity.

Very recently, another series of popular hypotheses seem to have successfully described the connection between the complexity on the CFT side and some geometric quantities of a black hole on the bulk side. In the theory of quantum information, complexity is a quantity that describes the minimal quantum gates needed to construct a quantum state or equally, how difficult a computational task would be. In order to calculate the complexity from the bulk, Susskind et. al. proposed three mutually connected hypotheses, which are complexity/length (CL) duality \cite{Stanford:2014jda}, complexity/volume (CV) duality \cite{Susskind:2014rva}, and complexity/action (CA) duality \cite{Brown:2015lvg,Brown:2015bva}. In these three hypotheses, the complexity respectively corresponds to the length of the Einstein-Rosen brige (ERB), the volume of ERB and the action of the Wheeler-DeWitt (WDW) patch, where the WDW patch is defined as the domain of the Cauchy surface anchored at the boundary state. Especially in the CA conjecture, the complexity $\mathcal{C}$ of a boundary state is related to the action $S$ of the WDW patch in the bulk,
\begin{equation}\label{CA}
\mathcal{C}=\frac{S}{\pi\hbar},
\end{equation}
where $\hbar$ is the reduced planck constant, and $\pi$ is a dimensionless constant \cite{Brown:2015lvg} which should not be mistaken for the circumference ratio.

In the literature, the complexity growth rates, or equivalently the action growth rates (AGRs) of the WDW patch for some typical black holes have been well studied \cite{Brown:2015lvg,Brown:2015bva,Cai:2016xho}. For the Schwarzschild-AdS black hole, the AGR is
\begin{equation}\label{Sch_AGR}
\frac{dS}{dt}=2M.
\end{equation}
For the Reissner-Nordstrom (RN)-AdS black hole, it is
\begin{equation}
\frac{dS}{dt}=[(M-\mu Q)_{+}-(M-\mu Q)_{-}].
\end{equation}
For rotating black holes such as the BTZ black hole and the Kerr-AdS black hole, they are both of the form
\begin{equation}
\frac{dS}{dt}=[(M-\Omega J)_{+}-(M-\Omega J)_{-}].
\end{equation}
The quantities with subscripts $\pm$ are evaluated on the outer/innner horizon.

So far there two popular schemes to calculate the AGR of the WDW patch in the literature. Brown, Roberts, Susskind, Swingle and Zhao (BRSSZ) \cite{Brown:2015lvg,Brown:2015bva} first achieved the correct result by directly integrating the bulk action inside the black hole and the boundary action on the horizon in the Einstein gravity. This scheme also turned out to be very convenient to calculate the AGR of black holes in other gravity theories \cite{Cai:2016xho,Huang:2016fks,Pan:2016ecg,Guo:2017rul,Abad:2017cgl}. Later in Refs. \cite{Lehner:2016vdi}, Lehner, Myers, Poisson and Sorkin (LMPS) pointed out that the boundary term, which is relegated to the York-Gibbons-Hawking (YGH) surface action in the BRSSZ scheme, is much trickier than one expected previously. In principle, the contributions from spacelike, timelike and null hypersurfaces and their corners should be computed separately. In practice, after cancellations in pairs, only double-null corner terms survive. In the LMPS scheme \cite{Lehner:2016vdi}, they summed up the bulk term, corner terms and surface terms after carefully studying the WDW patch in the Penrose diagram. This scheme was further developed in Refs. \cite{Jubb:2016qzt,Carmi:2016wjl,Reynolds:2016rvl}. Miraculously, in several typical examples, the two schemes yield exactly the same final results \cite{Lehner:2016vdi}. Following the LMPS scheme, lots of other examples of the AGR of black holes have been worked out, see Refs. \cite{Cai:2017sjv,Tao:2017fsy,Wang:2017uiw,Miao:2017quj,Jiang:2018sqj,Jiang:2018pfk,Meng:2018vtl,Brown:2018bms} as a partial list.

The Einstein gravity has been well tested in many circumstances. However, it remains yet to be verified in the non-linear, dynamically strong-curvature or strong-field regime. Among many modified gravity theories, the Chern-Simons (CS) modified gravity \cite{Jackiw:2003pm,Alexander:2009tp} stands out as one of the few examples whose Dirichlet boundary problem has been well studied \cite{Grumiller:2008ie}. By this advantage, we can study it explicitly in many scenarios, e.g. the membrane paradigm of black holes \cite{Saremi:2011ab,Fischler:2015kro,Zhao:2015inu}. Stimulated by the LIGO observational results \cite{Abbott:2018utx}, the CS modified gravity has attracted reviving interests recently \cite{Loutrel:2018ydv,Gao:2018acg,Tahura:2018zuq,Nishizawa:2018srh}.

In this work, we will calculate the AGRs of black holes in the CS modified gravity, including the Schwarzschild-AdS black hole and a slowly-rotating black hole solution discovered by Yagi \cite{Yagi:2012ya}. As a very quick warmup, we will recall the AGRs of the Schwarzschild-AdS and Kerr-AdS black holes in section \ref{subsect-Ein}, and collect the bulk and Dirichlet boundary terms of action in the CS modified gravity in section \ref{subsect-act-CS}. In section \ref{sect-sph}, the AGR of Schwarzschild-AdS black holes in CS modified gravity will be calculated in both the BRSSZ and the LMPS schemes. For the slowly rotating black hole \cite{Yagi:2012ya} in CS modified gravity, its AGR of will be calculated in section \ref{sect-rot} and confronted against the Schwarzschild and Kerr black holes in Einstein gravity. Further discussions and open problems will be presented in section \ref{sect-disc}.

Throughout this paper we will follow the convention of notations in Ref. \cite{Yagi:2012ya}, setting $G_N=\hbar=c=1$ and $\kappa_g=(16\pi)^{-1}$.

\section{Quick warmup}\label{sect-setup}
\subsection{AGR of black holes in Einstein gravity}\label{subsect-Ein}
The full action of a black hole consists of not only the bulk term but also the boundary term,
\begin{equation}
S=S_{\bk}+S_{\bd}.
\end{equation}
This is because the variation of the action should vanish according to the principle of least action. And it is natural that an extra boundary term would occur in the variation of action. To eliminate this unwanted term, the Dirichlet condition should be imposed on an appropriate boundary to address the problem.

In Einstein gravity with the cosmological constant $\Lambda=-3/L^2$,  the bulk term is the Einstein-Hilbert action
\begin{equation}
S_{\bk}=S_{EH}=\kappa_g\int d^4x\sqrt{-g}(R-2\Lambda),
\end{equation}
and the boundary term is the well-known YGH term
\begin{equation}
S_{\bd}=S_{YGH}=2\kappa_g\int_\Sigma d^3x\sqrt{|h|}K,
\end{equation}
where $K$ is the trace of the extrinsic curvature $K_{\mu\nu}$ of the 3-dimensional boundary $\Sigma$, and $h$ is the determinant of the induced metric on the boundary.

It has been well studied that for a $D$-dimensional Schwarzschild-AdS black hole, the AGR is precisely \cite{Brown:2015lvg,Brown:2015bva}
\begin{equation}
\frac{dS}{dt}=\left[\left(\frac{D-1}{D-2}\right)M+\frac{(D-2)\Omega_{D-2}r^{D-3}}{8\pi G}\left(1+\frac{r^2}{L^2}\right)\right]\Bigg|^{r_h}_0=2M,
\end{equation}
where $r_h$ is the event horizon radius of the Schwarzschild-AdS black hole. In the $L\rightarrow\infty$ limit, it takes the form
\begin{equation}
\frac{dS}{dt}=8\pi\kappa_g(2r-3M)\Big|^{r_h}_0=2M.
\end{equation}

Also, the result for Kerr-AdS black hole has been carried out in Ref. \cite{Cai:2016xho},
\begin{equation}
\frac{dS}{dt}=\frac{1}{4G\Xi}\int _{0} ^{\pi}d\theta \sin{\theta}\left(\frac{r\Delta}{\rho^2}+\frac{\Delta '(r)}{2}\right)\Bigg|_{r_{-}} ^{r_{+}},
\end{equation}
where $r_\pm$ stands for the radius of inner or outer horizon of the black hole. In this case, we have
\begin{equation}\label{Kerr_AGR}
\frac{dS}{dt}=\left.\frac{1}{2G}\left(\frac{\Delta}{a}\arctan\frac{a}{r}+r+\frac{r^3}{L^2}\right)\right|^{r_{+}}_{r_{-}}
\end{equation}
with the spin parameter $a=J/M$. In the limit $L\rightarrow\infty$, the Kerr-AdS metric reduces to
\begin{equation}\label{Kerr metric}
ds_{K}^2= -\left( 1-\frac{2Mr}{\Sigma} \right) dt^2 - \frac{4Mar\sin^2\theta}{\Sigma}dtd\phi + \frac{\Sigma}{\Delta} dr^2
 + \Sigma d\theta^2 + \left( r^2+a^2 + \frac{2Ma^2 r \sin^2\theta}{\Sigma}  \right) \sin^2\theta d\phi^2,
\end{equation}
while $\Delta$ and $\Sigma$ tend to
\begin{equation}
\Delta =r^2 - 2Mr + a^2 ,\ \
\Sigma =r^2 + a^2 \cos^2\theta.
\end{equation}

The above results are obtained earlier in Refs. \cite{Brown:2015lvg,Brown:2015bva,Cai:2016xho} using the BRSSZ scheme. Later in Ref. \cite{Lehner:2016vdi}, Lehner et. al. confirmed these results using the LMPS scheme. It is less appreciated that, in order to get the Schwarzschild limit of Eq. \eqref{Kerr_AGR}, one should take firstly $a=0$ and then $r_{-}=0$.

\subsection{Bulk and boundary terms in CS modified gravity}\label{subsect-act-CS}
In the CS modified Gravity, the bulk integral comprises the Einstein-Hilbert and the CS term \cite{Jackiw:2003pm},
\begin{eqnarray}
\nonumber S_{\bk}&=&S_{EH}+S_{CS}\\
&=&\kappa_g\int d^4x\sqrt{-g}(R-2\Lambda+\frac{1}{4}\thecs{^{\ast}RR}),
\end{eqnarray}
where the Chern-Pontryagin density is defined as
\begin{equation}
{^{\ast}RR}=R_{\nu\mu\rho\sigma}{^\ast R}^{\mu\nu\rho\sigma}=R_{\nu\mu\rho\sigma}\frac{1}{2}\epsilon^{\rho\sigma\alpha\beta}R^{\mu\nu}_{\  \  \  \  \alpha\beta}.
\end{equation}
The boundary term is given by the YGH term plus the Grumiller-Mann-McNees (GMM) term \cite{Grumiller:2008ie},
\begin{eqnarray}
\nonumber S_{\bd}&=&S_{GHY}+S_{bCS}\\
&=&2\kappa_g\int_\Sigma d^3x\sqrt{|h|}(K+\frac{1}{2}\thecs n_\mu\epsilon^{\mu\nu\rho\sigma}K_\nu^{\ \delta}\nabla_\rho K_{\sigma\delta}),
\end{eqnarray}
where $n_\mu$ is the normal vector near the event horizon, and $\thecs$ is the CS coupling.

For a spacelike normal vector $n_\mu$, which is taken to point outwards, the hypersurface $\Sigma$ is timelike. Then we define the induced metric
\begin{equation}\label{inducedmetric1}
h_{\mu\nu}=g_{\mu\nu}-n_\mu n_\nu
\end{equation}
as will be done in subsection \ref{EinsteinHi-GibbonHk_term}. Contrarily, as will be the case in section \ref{sect-rot}, a timelike normal vector $n_\mu$ in the boundary terms should point inwards \cite{Brown:2015lvg}, and the induced metric of the spacelike hypersurface $\Sigma$ is defined as
\begin{equation}\label{inducedmetric2}
h_{\mu\nu}=g_{\mu\nu}+n_\mu n_\nu.
\end{equation}

If one treats $\thecs$ as a dynamical field, the full action will also involve its kinetic and potential terms, see equation \eqref{actionfull}. In this paper, we will work in the non-dynamical framework in section \ref{sect-sph}, and dynamical in section \ref{sect-rot}.

\section{AGR of Schwarzschild black holes in CS modified gravity}\label{sect-sph}
Finding exact solutions of CS modified is difficult. Fortunately, Campbell et. al. first showed, in the context of string theory, both the Schwarzschild and the FRW spacetime can lead to an exact CS three-form where the modified field equations are not affected \cite{Campbell:1990fu}. Later in Ref. \cite{Jackiw:2003pm}, Jackiw and Pi proved that, with a canonical choice of the CS scalar, the Schwarzschild spacetime remains a solution of the non-dynamical modified theory. For an extensive discussion on exact vacuum solutions of the CS modified gravity, the reader can refer to Ref. \cite{Alexander:2009tp}.

It is straightforward to check that the metric of the Schwarzschild-AdS black hole
\begin{equation}
ds^2=-\left(1-\frac{2M}{r}+\frac{r^2}{L^2}\right)dt^2+\left(1-\frac{2M}{r}+\frac{r^2}{L^2}\right)^{-1}dr^2+r^2(d\theta^2+\sin^2\theta d\phi^2)
\end{equation}
satisfies the field equations of non-dynamical CS modified gravity. Therefore, in this section, we will work in the non-dynamical framework and calculate the AGR of the Schwarzschild-AdS black hole. Remind that the AGR involves not only the metric of spacetime but also the gravitational term in the action of gravity theory. Although the Schwarzschild-AdS black hole in CS modified gravity has the same form as that in the Einstein gravity, we should scrutinize potential contributions from additional terms $S_{CS}$ and $S_{bCS}$ in the action.

In the rest of this section, we will calculate the AGR of the Schwarzschild-AdS black holes in the CS modified gravity by applying two different schemes mentioned above. In subsection \ref{EinsteinHi-GibbonHk_term}, following the BRSSZ scheme, we will calculate the bulk and boundary terms and then add them together to get the total AGR. In subsection \ref{corner_Schwar}, we will use the LMPS scheme to compute the corner and surface terms instead of boundary terms. Note that the two schemes share the equivalent bulk terms. As we will see in the end, the two schemes yield exactly the same result.

\subsection{The BRSSZ scheme}\label{EinsteinHi-GibbonHk_term}
In this subsection, we will apply the BRSSZ scheme \cite{Brown:2015lvg,Brown:2015bva} to the Schwarzschild-AdS black hole in the CS modified gravity. Following this scheme, we will directly calculate the Einstein-Hilbert action, the Chern-Pontryagin density term, the YGH term and the GMM term separately, and then assemble them together. To keep some generality, we will start with a general spherical black hole characterized by two functions $f_1(r)$ and $f_2(r)$. But finally we will set them to the special form \eqref{Sch}, yielding the result for the Schwarzschild-AdS black hole.

In general, the metric of a spherically symmetric static black hole has the form
\begin{equation}\label{metric_sph}
ds^2=g_{\mu\nu}dx^\mu dx^\nu=-f_1(r)dt^2+f_2(r)dr^2+r^2(d\theta^2+\sin^2\theta d\phi^2).
\end{equation}
The location of horizon/horizons is dictated by $f_1(r)=0$. The radius of outer horizon will be represented by $r_{+}$. The radius of inner horizon, if it exists, will be denoted by $r_{-}$. Corresponding to the above metric, the Ricci scalar is
\begin{equation}
R=\frac{2}{r^2}-\frac{2}{f_2r^2}-\frac{2f_1'}{f_1f_2r}+\frac{f_1'^2}{2f_1^2f_2}+\frac{2f_2'}{f_2^2r}+\frac{f_1'f_2'}{2f_1f_2^2}
-\frac{f_1''}{f_1f_2}.
\end{equation}
On each timelike hypersurface with constant $r$ near the horizon $r=r_\pm$, the induced metric can be defined as Eq. \eqref{inducedmetric1}, where the spacelike normal vector $n^\mu=\left(0,\sqrt{1/f_2},0,0\right)$. We can thus easily calculate the determinant of induced metric
\begin{equation}
\sqrt{-h}=r^2\sin{\theta\sqrt{f_1}}
\end{equation}
and the trace of extrinsic curvature
\begin{equation}
K=\nabla^\mu n_\mu=\frac{1}{\sqrt{-g}}\partial_\mu(\sqrt{-g}n^\mu)=\frac{2}{r\sqrt{f_2}}+\frac{f_1'}{2f_1\sqrt{f_2}}.
\end{equation}

Armed with the above geometric setup, we are now ready to compute the AGR of the Schwarzschild-AdS black hole. The Einstein-Hilbert term is
\begin{align}\label{Action_EH}
\nonumber \delta S_{EH}=&\kappa_g\delta t\int drd\theta d\phi\sqrt{-g}(R-2\Lambda) \\
=&2\pi\kappa_g\delta t\Bigg[\int^{r_+}_{r_-} dr\Bigg(4\sqrt{f_1f_2}-4\sqrt{\frac{f_1}{f_2}}-\frac{4rf_1'}{\sqrt{f_1f_2}}+\frac{r^2f_1'^2}{f_1\sqrt{f_1f_2}}
+\frac{4 rf_2'\sqrt{f_1f_2}}{f_2^2}+\frac{r^2f_1'f_2'}{f_2\sqrt{f_1f_2}}-\frac{2r^2f_1''}{\sqrt{f_1f_2}}+\frac{6}{L^2}\Bigg).
\end{align}
One can also find out that for the spherical spacetime \eqref{metric_sph}, the Chern-Pontryagin density vanishes, $^{\ast}RR=R_{\nu\mu\rho\sigma}\frac{1}{2}\epsilon^{\rho\sigma\alpha\beta}R^{\mu\nu}_{\  \  \  \  \alpha\beta}=0$. That means the CS term in the action is simply zero,
\begin{equation}
\delta S_{CS}=\kappa_g\delta t\int drd\theta d\phi\frac{1}{4}\thecs{^{\ast}RR}=0.
\end{equation}

On the boundary, the YGH term is
\begin{align}
\nonumber \delta S_{YGH}=&2\kappa_g\delta t\int_\Sigma d\theta d\phi\sqrt{-h}K \Bigg| ^{r_{+}}_{r_{-}}\\
=&4\pi\kappa_g\delta t \left[\frac{r(4f_1+rf_1')}{\sqrt{f_1f_2}}\right]\Bigg| ^{r_{+}}_{r_{-}}.
\end{align}
Just like the Chern-Pontryagin density, the boundary integrand of CS modified gravity $n_\mu\epsilon^{\mu\nu\rho\sigma}K_\nu^{\ \delta}\nabla_\rho K_{\sigma\delta}$ is also trivial, which again gives us another vanishing term in the action,
\begin{equation}
\delta S_{bCS}=2\kappa_g\int_\Sigma d^3x\sqrt{-h} \frac{1}{2}\thecs n_\mu\epsilon^{\mu\nu\rho\sigma}K_\nu^{\ \delta}\nabla_\rho K_{\sigma\delta}=0.
\end{equation}

At this point, we can assemble them together. The bulk term is
\begin{align}\label{Action_bulk}
\nonumber \delta S_{\bk}=&\delta S_{EH}+\delta S_{CS} \\
=&2\pi\kappa_g\delta t\Bigg[\int^{r_+}_{r_-} dr\Bigg(4\sqrt{f_1f_2}-4\sqrt{\frac{f_1}{f_2}}-\frac{4rf_1'}{\sqrt{f_1f_2}}+\frac{r^2f_1'^2}{f_1\sqrt{f_1f_2}}
+\frac{4 rf_2'\sqrt{f_1f_2}}{f_2^2}+\frac{r^2f_1'f_2'}{f_2\sqrt{f_1f_2}}-\frac{2r^2f_1''}{\sqrt{f_1f_2}}+\frac{6}{L^2}\Bigg),
\end{align}
while the boundary term is naively
\begin{align}
\nonumber \delta S_{\bd}=&\delta S_{YGH}+\delta S_{bCS} \\
=&4\pi\kappa_g\delta t \left[\frac{r(4f_1+rf_1')}{\sqrt{f_1f_2}}\right]\Bigg| ^{r_{+}}_{r_{-}}.
\end{align}
After summation and simplification, we obtain the total AGR
\begin{eqnarray}
\nonumber\frac{dS}{dt}&=&\nonumber\frac{dS_{\bk}}{dt}+\frac{dS_{\bd}}{dt} \\
&=&8\pi\kappa_g\int^{r_+}_{r_-}dr\Bigg(\sqrt{f_1f_2}+\sqrt{\frac{f_1}{f_2}}+\frac{4r}{\sqrt{f_1}}+\frac{rf_2'\sqrt{f_1f_2}}{f_2^2}-\frac{f_1}{f_2\sqrt{f_2}}+\frac{6}{L^2}\Bigg).
\end{eqnarray}

For Schwarzschild-AdS black holes,
\begin{align}\label{Sch}
f_1=\frac{1}{f_2}=\left(1-\frac{2M}{r}+\frac{r^2}{L^2}\right),
\end{align}
the above expression of AGR reduces to
\begin{equation}
\frac{dS}{dt}=2M,
\end{equation}
coinciding with the result in the Einstein gravity.

The conclusion can be made more general. We have demonstrated that, both $\delta S_{CS}$ and $\delta S_{bCS}$ vanish for the metric \eqref{metric_sph}. In other words, compared with the Einstein gravity, the potential new contributions to the AGR in the CS modified gravity vanish for the bulk term and the boundary term separately. This result is robust for any spherically symmetrical spacetime in non-dynamical CS modified gravity, including the Reissner-Nordstr\"{o}m black hole, etc.

\subsection{The LMPS scheme}\label{corner_Schwar}
In the Einstein gravity, as pointed out in Ref. \cite{Lehner:2016vdi} and refined in Refs. \cite{Jubb:2016qzt,Carmi:2016wjl,Reynolds:2016rvl}, the contribution of corner terms of the null joints on the boundary of WDW patch should also be considered when calculating the AGR due to the time evolution. The same logic ought to be valid for other gravity theories. Following this logic, in a recent paper \cite{Jiang:2018sqj}, Jiang and Zhang systematically derived the corner terms and surface terms on any non-smooth boundary in the $f(R_{abcd},g_{ab})$ gravity. Fortunately, their results are applicable to the CS modified gravity considered in the present paper. In this subsection, we will follow Refs. \cite{Lehner:2016vdi,Jubb:2016qzt,Carmi:2016wjl,Reynolds:2016rvl,Jiang:2018sqj} and calculate the corner term as well as the surface term of Schwarzschild-AdS black holes in the WDW patch.

Before going to the Schwarzschild-AdS black holes, let us consider a subcase of the metric \eqref{metric_sph}, in which $1/f_2=f_1\equiv f(r)$, that is
\begin{equation}\label{metric_schw}
ds^2=-f(r)dt^2+\frac{1}{f(r)}dr^2+r^2d\theta^2+r^2\sin{\theta}^2d\phi^2.
\end{equation}
For simplicity, we will focus on the situation with a single horizon, which means equation $f(r)=0$ has only one positive root $r=r_{+}$.

\begin{figure}
\centering
\includegraphics[width=0.4\textwidth]{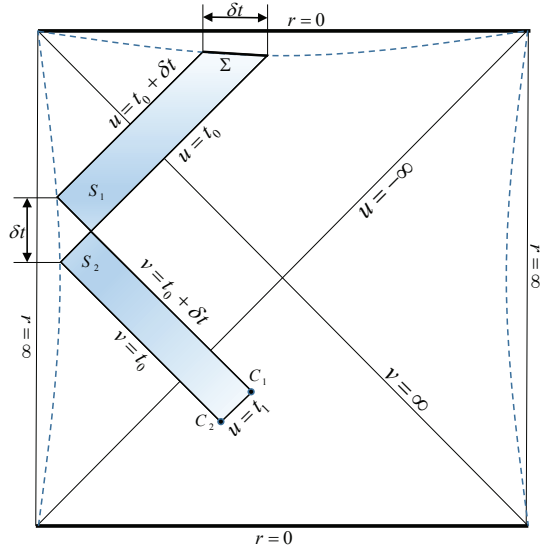}\\
\caption{The WDW patch of a Schwarzschild-AdS black hole. As illustrated, as time evolves from $t_0$ to $t_0+\delta t$, the action growth is given by $S_1$ and $S_2$ in two subregions. They are bounded by several null boundaries such as $v=t_0$, $v=t_0+\delta t$, $u=t_0$, $u=t_0+\delta t$, $u=t_1$ and a spacelike boundary $\Sigma$, where $t_1$ is the time coordinate of $C_1$ and $C_2$. As shown by the blue dashed curves, we should apply a cut-off $r=r_{\text{min}}$ near the singularity and $r=r_{\text{max}}$ near the asymptotic AdS boundaries.}\label{Penrose}
\end{figure}

The Penrose diagram of the WDW patch of spherically symmetric black hole with a single horizon (Schwarzschild-AdS black hole) is shown in Fig. \ref{Penrose}. During the time interval $[t_0, t_0+\delta t]$, the growth of action can be expressed as $\delta S=S(t_0 +\delta t,t_1)-S(t_0,t_1)$. As we can see in this figure, the corner terms arise from two double-null joints located at $r=r_1$ and $r=r_2$ which are denoted by $C_1$ and $C_2$ respectively.

For the subcase $1/f_2=f_1=f(r)$, the bulk term in the LMPS scheme has a form similar to Eq. \eqref{Action_bulk},
\begin{equation}
\delta S_{\bk}=-4\pi\kappa_g r\left(2f+rf'\right)\delta t \Big| _{r=r_1}.
\end{equation}

In order to derive the corner term and the surface term, we have to rearrange the bulk action in the form
\begin{align}\label{ac_general}
\nonumber S_{\bk}=&\kappa_g \int d^4 x \sqrt{-g} \mathcal{L}(R_{abcd},g_{ab}) \\
=&\kappa_g \int d^4 x \sqrt{-g} [\mathcal{L}(\varphi_{abcd},g_{ab})-\psi^{abcd}(\varphi_{abcd}-R_{abcd})]
\end{align}
and work out the auxiliary fields $\psi_{abcd}$, $\varphi_{abcd}$. Here $\psi_{abcd}$ and $\varphi_{abcd}$ have the same symmetries in indices as the Riemann tensor $R_{abcd}$, while the Lagrangian density of the CS modified gravity is
\begin{equation}
\mathcal{L}(R_{abcd},g_{ab})=R-2\Lambda+\frac{1}{4}\thecs R_{\nu\mu\rho\sigma}\frac{1}{2}\epsilon^{\rho\sigma\alpha\beta}R^{\mu\nu}_{\  \  \  \  \alpha\beta}.
\end{equation}
The variation of action \eqref{ac_general} is
\begin{equation}
\delta S_{\bk}=\kappa_g \int d^4 x \sqrt{-g}(E_{ab}\delta g^{ab}+E^{abcd}_\varphi\delta \varphi_{abcd}+E^{abcd}_{\psi}\delta \psi_{abcd})-\delta S_{\bd}.
\end{equation}
Recalling Eq. (2.7) in Ref. \cite{Jiang:2018sqj},
\begin{equation}
E_{\varphi}^{abcd}=\frac{\partial\mathcal{L}(\varphi_{abcd},g_{ab})}{\partial\varphi_{abcd}}-\psi ^{abcd},\ \ \ \ \
E_{\psi}^{abcd}=R^{abcd}-\varphi^{abcd},
\end{equation}
after imposing the equations of motion $E_{\psi}^{abcd}=0$ and $E_{\varphi}^{abcd}=0$, we find the two auxiliary fields take the form
\begin{equation}\label{pshi}
\psi_{abcd}=\frac{\partial\mathcal{L}(\varphi_{abcd},g_{ab})}{\partial\varphi_{abcd}},~~~~\varphi_{abcd}=R_{abcd}.
\end{equation}
For the CS modified gravity, the auxiliary field $\psi^{abcd}$ in Eq. \eqref{pshi} can be calculated straightforwardly,
\begin{equation}\label{psiabcd}
\psi^{abcd}=\frac{1}{2}\left[(g^{ac}g^{bd}-g^{ad}g^{bc})+\frac{1}{4}\thecs\epsilon^{cd\rho\sigma}(g^{\mu b}g^{\nu a}-g^{\mu a}g^{\nu b})R_{\mu\nu\rho\sigma}\right].
\end{equation}

To proceed, we should also calculate the scalar
\begin{align}\label{psihat_ori}
\nonumber \hat{\Psi}=&4\psi_{abcd}k^a l^b k^c l^d \\
=&-2+\frac{1}{4}\thecs\epsilon^{cd\rho\sigma}k^{\nu}k^{\mu}k_{\rho}k_{\sigma}R_{\mu\nu\rho\sigma}
\end{align}
where $k^a$ and $l^a$ are two null vectors. They are respectively normal to and along the null boundary of the WDW patch, satisfying $l_a l^a=0$ and $k_a l^a=-1$. It can be checked that the second term in Eq. \eqref{psihat_ori} is actually vanished for the metric \eqref{metric_schw}, which means
\begin{equation}\label{psihat}
\hat{\Psi}=-2.
\end{equation}

To follow the convention of \cite{Jiang:2018sqj}, it is convenient to choose $k_{1a}=\partial_a u$ and $k_{2a}=\partial_a v$, where $u=t+r^*(r)$ and $v=t-r^*(r)$ is the ingoing and outgoing Eddington coordinate with $r^*=\int f^{-1} dr$. In fact, $k_1$ is the normal vector of hypersurfaces $u=\const$, and $k_1$ is the normal vector of hypersurfaces $v=\const$. With this choice, we have
\begin{equation}\label{k1k2}
k_1 \cdot k_2=\frac{2}{f}.
\end{equation}
As shown in Fig. \ref{Penrose}, the evolution of the WDW patch involves five null hypersurfaces $u=t_1$, $u=t_0$, $u=t_0+\delta t$, $v=t_0$, $v=t_0+\delta t$. Their intersections give rise to the so-called corner terms in the full action. Following Ref. \cite{Jiang:2018sqj}, the corner term of WDW patch is
\begin{equation}\label{actioncorner}
S_{\cn}=\int d^4 x \hat{\Psi} \log{\left(-\frac{1}{2}k_1 \cdot k_2\right)}.
\end{equation}
Most of such terms have cancelled with each other. By inserting \eqref{psihat}, \eqref{k1k2} into \eqref{actioncorner}, and taking the difference between corner terms at $C_1$ and $C_2$, we get the net contribution from the corner terms
\begin{equation}
\delta S_{C_1}-\delta S_{C_2}=4\pi\kappa_g\delta t\left[r^2f'+2rf\log{(-f)}\right] \Bigg| _{r=r_1}.
\end{equation}
Here we have used $\delta r=r_1-r_2=-\frac{1}{2}f(r_1) \delta t$.

According to Ref. \cite{Jiang:2018sqj}, the surface term can be written as
\begin{align}
\nonumber\delta S_{\Sigma}=&\kappa_g \delta t \hat{\Psi}\int_{\Sigma}K d\Sigma \\
=&4\pi\kappa_g\delta t \left[r(4f+rf')\right]\Big| _{r=0},
\end{align}
where we have applied the cut-off $r_{\textrm{min}}\rightarrow 0$.

Specifically, for the Schwarzschild-AdS spacetime
\begin{equation}
f(r)=\left(1-\frac{2M}{r}+\frac{r^2}{L^2}\right),
\end{equation}
we can compute the bulk term, the corner term and the surface term, and then sum them up to get the AGR
\begin{align}
\nonumber\frac{dS}{dt}=&\frac{d}{dt}\left(S_{\bk}+S_{\Sigma}+S_{\cn}\right) \\
\nonumber=&4\pi\kappa_g \left[-\frac{2r^2}{L^2}+6M+\left(2M+\frac{2r^2}{L^2}\right)\right]\Bigg| _{r=r_1}\\
=&2M.
\end{align}

This result coincides with the final equation in section \ref{EinsteinHi-GibbonHk_term}. This does not surprise us, because in all worked examples in the Einstein gravity, the BRSSZ scheme and the LMPS scheme yield exactly the same final results \cite{Lehner:2016vdi}.

\section{AGR of slowly rotating CS black holes}\label{sect-rot}
So far we have discussed the AGR of spherically symmetric solutions in the non-dynamical CS modified gravity. Boringly, in all these examples, the AGR is not modified by the CS term and related terms. So we would like to turn to a nontrivial example. A good candidate is the slowly rotating black hole in the dynamical CS modified gravity. In Ref. \cite{Yagi:2012ya}, such a black hole has been derived perturbatively to the quadratic order in the angular momentum. In section \ref{subsect-Yagi}, we will briefly review this black hole solution in the dynamical CS modified gravity. In section \ref{subsect-rot}, we will utilize the BRSSZ scheme to compute its AGR. The final result will be different from but consistent with that of the Kerr-AdS black holes in the Einstein gravity.

\subsection{Brief review of slowly rotating CS black holes}\label{subsect-Yagi}
From now on, we shift from the non-dynamical CS modified gravity to the dynamical one. The action of the dynamical CS modified gravity is slightly more complicated than that in subsection \ref{subsect-act-CS},
\begin{align}\label{actionfull}
\nonumber S=& \int d^4x\sqrt{-g} \left\{ \kappa_g R+\frac{\alpha}{4}\thecs\ {}^{\ast}RR-\frac{\beta}{2}[\nabla_\mu\thecs\nabla^{\mu}\thecs + 2 V(\thecs)]\right\}  \\
& + \int d^3x\sqrt{|h|}\left(2\kappa_g K+\frac{\alpha}{2}\thecs n_\mu\epsilon^{\mu\nu\rho\sigma}K_\nu^{\ \delta}\nabla_\rho K_{\sigma\delta} \right)\Bigg| ^{r_{+}}_{r_{-}},
\end{align}
where $\thecs$ is a scalar field and $\alpha$, $\beta$ are two coupling constants. In this paper, it suffices to restrict our discussion to the case $V(\thecs)=0$, which is correct for both spherical black holes \cite{Alexander:2009tp} and slowly rotating black holes \cite{Yagi:2012ya}. By making $\thecs$ and $\beta$ dimensionless and $\alpha$ to have the dimension of $(\textrm{length})^2$, the solution of the scalar field is \cite{Yagi:2012ya}\footnote{In this paper, we trust on the print of the most recent version arXiv:1206.6130v4 [gr-qc] of Ref. \cite{Yagi:2012ya}.}
\begin{equation}\label{thecs}
\thecs\approx\frac{5\alpha\chi\cos{\theta}}{8\beta r^2}\left(1+\frac{2M}{r}+\frac{18M^2}{5r^2}\right),
\end{equation}
where $M$ is the mass of the black hole and $\chi=a/M$ is a dimensionless spin parameter.

In terms of another dimensionless parameter $\zeta=\alpha^2/(\kappa_g\beta M^4)$, to the second order of $\chi$, the metric of a slowly rotating black hole in the CS modified gravity can be written as \cite{Yagi:2012ya}
\begin{equation}\label{KerrCS metric}
ds^{2}=ds_K^2+ 2 g_{t\phi}^{CS} dtd\phi + g_{tt}^{CS} dt^2 + g_{rr}^{CS} dr^2 + g_{\theta \theta}^{CS} d\theta^2 + g_{\phi\phi}^{CS} d\phi^2.
\end{equation}
The Kerr metric $ds_K^2$ is given by Eq. \eqref{Kerr metric}, and the other components are given by
\begin{align}
g_{t\phi}^{CS}=&\frac{5}{8}\zeta\chi\frac{M^5}{r^4}\left(1+\frac{12M}{7r}+\frac{27M^2}{10r^2}\right)\sin^2{\theta}+\mathcal{O}(\alpha^2\chi^3),\\
g_{tt}^{CS}=&\zeta\chi^2\frac{M^3}{r^3}\Bigg[\frac{201}{1792}\left(1+\frac{M}{r}+\frac{4474}{4221}\frac{M^2}{r^2}-\frac{2060}{469}\frac{M^3}{r^3}+\frac{1500}{469}\frac{M^4}{r^4}-\frac{2140}{201}\frac{M^5}{r^5}+\frac{9256}{201}\frac{M^6}{r^6}-\frac{5376}{67}\frac{M^7}{r^7}\right)\nonumber\\
&\times(3\cos^2\theta-1)-\frac{5}{384}\frac{M^2}{r^2}\left(1+100\frac{M}{r}+194\frac{M^2}{r^2}+\frac{2220}{7}\frac{M^3}{r^3}-\frac{1512}{5}\frac{M^4}{r^4}\right)\Bigg]+\mathcal{O}(\alpha^2\chi^4),\\
g_{rr}^{CS}=&\zeta\chi^2\frac{M^3}{r^3f(r)^2}\Bigg[\frac{201}{1792}f(r)\left(1+\frac{1459}{603}\frac{M}{r}+\frac{20000}{4221}\frac{M^2}{r^2}+\frac{51580}{1407}\frac{M^3}{r^3}-\frac{7580}{201}\frac{M^4}{r^4}-\frac{22492}{201}\frac{M^5}{r^5}-\frac{40320}{67}\frac{M^6}{r^6}\right)\nonumber\\
&\times(3\cos^2\theta-1)-\frac{25}{384}\frac{M}{r}\left(1+3\frac{M}{r}+\frac{322}{5}\frac{M^2}{r^2}+\frac{198}{5}\frac{M^3}{r^3}+\frac{6276}{175}\frac{M^4}{r^4}-\frac{17496}{25}\frac{M^5}{r^5}\right)\Bigg]+\mathcal{O}(\alpha^2\chi^4),\\
g_{\theta\theta}^{CS}=&\frac{201}{1792}\zeta\chi^2M^2\frac{M}{r}\left(1+\frac{1420}{603}\frac{M}{r}+\frac{18908}{4221}\frac{M^2}{r^2}+\frac{1480}{603}\frac{M^3}{r^3}+\frac{22460}{1407}\frac{M^4}{r^4}+\frac{3848}{201}\frac{M^5}{r^5}+\frac{5376}{67}\frac{M^6}{r^6}\right)\nonumber\\
&\times(3\cos^2\theta-1)+\mathcal{O}(\alpha^2\chi^4),\\
g_{\phi\phi}^{CS}=&\sin^2\theta g_{\theta\theta}^{CS}+\mathcal{O}(\alpha^2\chi^4).
\end{align}
Since $\zeta=\alpha^2/(\kappa_g\beta M^4)$, it is easy to see that the presented $CS$ corrections are of $\mathcal{O}(\alpha^2\chi^2)$. It is worth noting that, as a result, this metric goes back to the Schwarzschild metric in the case $\chi=0$, and to the Kerr metric in the case $\alpha=0$.

The location of the outer/inner horizon $r=r_{\pm}$ can be found by solving the equation $g_{tt}g_{\phi\phi}-g_{t\phi}^2=0$. The approximate solutions are \cite{Yagi:2012ya}
\begin{align}
r_{+}&=r_{+,K}-\frac{915}{28672}\zeta\chi^2M+\mathcal{O}\left(\alpha^2\chi^4\right),\label{rp}\\
r_{-}&=r_{-,K}+\frac{18432\zeta M(3\cos^2{\theta}-1)}{\chi^{16}(4\cos^2{\theta}+\chi^2)}+\mathcal{O}\left(\alpha^2\chi^{-14}\right).\label{rm}
\end{align}
where $r_{+,K}$ and $r_{-,K}$ denote the radii of outer and inner horizons of the Kerr black hole with the same mass and spin. To guarantee the existence of inner horizon, the second term and higher order terms in Eq. \eqref{rm} should be suppressed. This imposes the condition $\zeta\lesssim\mathcal{O}(\chi^{20})$.

\subsection{Action Growth Rate}\label{subsect-rot}
In this subsection, we will implement the BRSSZ scheme to compute the growth rate of the action \eqref{actionfull} for the metric \eqref{KerrCS metric}. The procedure is parallel to Section \ref{EinsteinHi-GibbonHk_term}, except for that we now have to take into account the kinetic term of scalar field $\thecs$. In the final expression of AGR, we will neglect terms of $\mathcal{O}(\chi^6)$ and $\alpha^2\mathcal{O}(\chi^4)$, etc. Since the metric \eqref{KerrCS metric} makes sense only in the limit of small $\alpha$ and small $\chi$, it is legitimate for us to neglect high-order terms. In accordance with this approximation, we will only present terms of appropriate orders in the intermediate results such as the Ricci scalar, the Chern-Pontryagin density, and so on.

Starting with the metric \eqref{KerrCS metric}, we can calculate the Ricci scalar
\begin{align}
\nonumber R=&-\frac{\alpha^2\chi^2}{256\beta\kappa_g r^{11}}
\Big[10368M^5+3132M^4r-1420M^2r^3-500Mr^4-125r^5 \\
\nonumber & +(10368M^5+3780M^4r+720M^3r^2-860Mr^3-300Mr^4-75r^5)\cos{2\theta}\Big] \\
&+\alpha^2\mathcal{O}(\chi^4)+\alpha^4\mathcal{O}(1)+\cdots,
\end{align}
in which the lowest order term is of $\mathcal{O}(\alpha^2\chi^2)$, and the Chern-Pontryagin density
\begin{align}
^{\ast}RR= -\frac{144\chi  M^3\sin{2\theta}}{r^5}+\mathcal{O}(\chi^3)+\alpha^2\mathcal{O}(\chi)+\alpha^4\mathcal{O}(1)+\cdots.
\end{align}
The kinetic term of scalar field $\thecs$ is
\begin{align}
\nonumber \nabla_\mu\thecs\nabla^{\mu}\thecs=& \frac{\alpha^2\chi^2}{128\beta r^{11}}\Big[4(2M-r)(36M^2+15Mr+5r^2)^2\cos^2{\theta}-(18M^2+10Mr+5r^2)^2r\sin^2{\theta}\Big] \\
&+\alpha^2\mathcal{O}(\chi^4)+\alpha^4\mathcal{O}(1)+\cdots.
\end{align}
In this subsection, ellipses denote the higher order terms in every equation.

In the current case, we introduce a spacelike hypersurface slightly inside the outer horizon and a hypersurface slightly outside the inner horizon. They are normal to the vector $n^\mu=\left(0,-1/\sqrt{-g_{rr}},0,0\right)$ pointing inwards. On each hypersurface, the induced metric has the determinant
\begin{equation}
\sqrt{h}=\sqrt{(2M-r)r^3}\sin{\theta} +\frac{\chi^2 M^2}{4(2M-r)^{1/2}r^{1/2}}\Big[2M-3r+(2 M-r)\cos{2 \theta}\Big]\sin{\theta}+\mathcal{O}(\chi^4)+\alpha^2\mathcal{O}(\chi^2)+\alpha^4\mathcal{O}(1)+\cdots,
\end{equation}
the GMM term is proportional to
\begin{align}
\sqrt{h}n_\mu\epsilon^{\mu\nu\rho\sigma}K_\nu^{\ \delta}\nabla_\rho K_{\sigma\delta}=\frac{3\chi M^2 (3M-r)}{4r^2} (\cos{\theta}-\cos{3\theta})+\mathcal{O}(\chi^3)+\alpha^2\mathcal{O}(\chi)+\alpha^4\mathcal{O}(1)+\cdots,
\end{align}
and the trace of the extrinsic curvature is given by
\begin{align}
\nonumber \sqrt{h}K=&(2r-3M)\sin{\theta}\\
\nonumber&+\frac{\chi^2 M^2}{4r^2}\Big[(2M+3r)\sin{\theta}+(2M-r)\sin{3\theta}\Big]\\
\nonumber&-\frac{\chi^4 M^4}{16r^4}\Big[2(2M+r)\sin{\theta}+(6M+r)\sin{3\theta}+(2M-r)\sin{5\theta}\Big]\\
\nonumber&+\frac{\alpha^2\chi^2}{150528\beta\kappa_g M r^8}\Big[804384M^6+887304M^5r+576300M^4r^2+150632M^3r^3\\
\nonumber&+34036M^4r^4+266Mr^5-3178Mr^5-12663r^6+9(606816M^6+150616M^5r\\
\nonumber&+82340M^4r^2-37336M^3r^3+892M^2r^4-3178Mr^5-4221r^6)\cos{2\theta}\Big] \\
&+\mathcal{O}(\chi^6)+\alpha^2\mathcal{O}(\chi^4)+\alpha^4\mathcal{O}(1)+\cdots.
\end{align}

After a lengthy but straightforward computation, we find the AGR is nontrivial
\begin{align}\label{CS-order2}
\nonumber\frac{dS}{dt}=&\Bigg\{8\pi\kappa_g(2r-3M)+\frac{16\chi^2M^2\pi\kappa_g(M+r)}{3r^2}-\frac{16\chi^4M^4\pi\kappa_g(3M+r)}{15r^4}+\mathcal{O}(\chi^6)\\
\nonumber&+\frac{\alpha^2\chi^2\pi}{48\beta r^8}\Bigg[-3888 M^5+288 M^4 r+360 M^3 r^2+670 M^2 r^3+80 M r^4+25 r^5\\
\nonumber&+12 M^2 (3 M-r) \left(18 M^2+10 M r+5 r^2\right) \left(\frac{r}{2 M-r}\right)^{1/2}\Bigg]\\
&+\alpha^2\mathcal{O}(\chi^4)+\alpha^4\mathcal{O}(1)+\cdots\Bigg\}\Bigg| ^{r_{+}}_{r_{-}}.
\end{align}
It is the main result of this section. To check and understand this result, it is helpful to incarnate it in two special cases. One is the case with $\chi=0$, and the other $\alpha=0$.

In the special case $\chi=0$, the metric \eqref{KerrCS metric} reduces to the Schwarzschild solution. Setting $\chi=0$ in Eq. \eqref{thecs} and inserting it into the action \eqref{actionfull}, we arrive at the Einstein-Hilbert action supplemented with the YGH term. Therefore, this case corresponds to the Schwarzschild black hole in the Einstein gravity, and the AGR should be confronted with Eq. \eqref{Sch_AGR}. Indeed, as we take firstly $\chi=0$ and secondly $r_{-}=0$, the expression \eqref{CS-order2} is consistent with Eq. \eqref{Sch_AGR}.

In the other case $\alpha=0$, the metric \eqref{KerrCS metric} reduces to the Kerr solution, and the action \eqref{actionfull} goes back to the Einstein gravity. In this special case, the AGR \eqref{CS-order2} is simplified to the form
\begin{align}\label{Kerr-CS_AGR2}
\frac{dS}{dt}=\Bigg[r+\frac{\chi^2M^2(M+r)}{3r^2}-\frac{\chi^4M^4(3M+r)}{15r^4}+\mathcal{O}(\chi^6)\Bigg]\Bigg| ^{r_{+}}_{r_{-}}.
\end{align}
For comparison, let us take a closer look at the AGR of slowly rotating Kerr black holes in the Einstein gravity. Making use of the formula
\begin{eqnarray}
\nonumber\arctan\frac{a}{r}&=&\sum_{n=0}^{\infty}\frac{(-1)^n}{2n+1}\left(\frac{a}{r}\right)^{2n+1}\\
&\approx&\frac{a}{r}-\frac{1}{3}\left(\frac{a}{r}\right)^3+\frac{1}{5}\left(\frac{a}{r}\right)^5,
\end{eqnarray}
we are able to expand Eq. \eqref{Kerr_AGR} in the slowly rotating limit as
\begin{align}\label{Kerr_AGR2}
\nonumber\frac{dS}{dt}\approx&\left.\frac{1}{2}\left\{\left(\frac{r^2}{a}-2M\frac{r}{a}+a\right)
\left[\frac{a}{r}-\frac{1}{3}\left(\frac{a}{r}\right)^3+\frac{1}{5}\left(\frac{a}{r}\right)^5\right]
+r\right\}\right|^{r_{+}}_{r_{-}} \\
\approx&\left[r+\frac{M+r}{3}\left(\frac{a}{r}\right)^2-\frac{3M+r}{15}\left(\frac{a}{r}\right)^4\right] \Bigg|^{r_{+}}_{r_{-}},
\end{align}
where we have omitted higher order terms and set the cosmological constant to zero. This equation represents the AGR of slowly rotating Kerr black holes in the Einstein gravity. Recall that Eq. \eqref{Kerr-CS_AGR2} is the $\alpha=0$ limit of the AGR of slowly rotating black holes in the CS modified gravity. They agree with each other very well.

\section{Discussion}\label{sect-disc}
In this paper, we have calculated the AGRs of the WDW patches for the Schwarzschild-AdS black hole in the non-dynamical CS modified gravity and the slowly rotating black hole \cite{Yagi:2012ya} in the dynamical CS modified gravity. The CS modified gravitational action is supplemented with the YGH and GMM boundary terms \cite{Grumiller:2008ie}. The AGR of the Schwarzschild-AdS black hole was calculated in both the BRSSZ scheme \cite{Brown:2015lvg,Brown:2015bva} and the LMPS scheme \cite{Lehner:2016vdi}, giving the same result $dS/dt=2M$. The AGR of the slowly rotating black hole was calculated only in the BRSSZ scheme \cite{Brown:2015lvg,Brown:2015bva}, yielding the result \eqref{CS-order2}. Our study is motivated by the CA duality conjecture \cite{Brown:2015lvg,Brown:2015bva} Eq. \eqref{CA}, which relates the AGRs to the growth rates of quantum complexity of s`tates in the dual field theory after the scrambling but before the recurrence. In connection with our results, there are three open problems left as follows.

First, when different gravity theories share the same black hole solution, we are not sure whether the AGRs are the same. In both the CL \cite{Stanford:2014jda} and CV \cite{Susskind:2014rva} duality conjectures, the holographic complexity is determined soly by the metric of spacetime. In the CA duality conjecture \cite{Brown:2015lvg,Brown:2015bva}, the AGR involves not only the metric of spacetime but also the action of gravity theory. Therefore, in principle, the CA duality may lead to different holographic complexities for the same solution in different gravitational theories. This provides us with an interesting diagnostic test for the CA duality conjecture. Our study shows that, for the Schwarzschild-AdS black holes in the Einstein gravity and the CS modified gravity, the AGRs are the same and the CA duality passed the test. However, it remains to be seen if the CA duality can always pass this diagnostic test in different cases.

Second, it is unclear if the sufferings in taking the Schwarzschild limit of the AGRs indicate some inconsistency of the CA duality in its present formulation. In the main text, we advocated taking the Schwarzschild limit of the AGR of rotating black holes by setting firstly $\chi=0$ and then $r_{-}=0$. Instead, if one had expressed $r_{-}$ in terms of the spin parameter, inserted it into Eq. \eqref{Kerr_AGR} or \eqref{CS-order2} and then taken $\chi=0$, one would have failed in achieving the Schwarzschild limit. Especially, Eqs. \eqref{CS-order2} and \eqref{Kerr_AGR2} would have been divergent.

Third, it is important to find the exact solution of rotating black holes in the CS modified gravity with a cosmological constant. Our study of the rotating black holes has been restricted to the small-spin regime without a cosmological constant. In the first line of Eq. \eqref{CS-order2}, which is sourced by $ds_K^2$ in Eq. \eqref{KerrCS metric}, we have deliberately omitted terms higher than $\mathcal{O}(\chi^4)$. It is expectable that this line can be replaced by the right hand side of Eq. \eqref{Kerr_AGR}, and, for the exact rotating solution to the CS modified gravity, the full form of Eq. \eqref{CS-order2} will no longer be divergent in the aforementioned limit.

\begin{acknowledgments}
This work is supported by the National Natural Science Foundation of China (Grant No. 91536218).
\end{acknowledgments}


\begin{thebibliography}{99}
\bibitem{Ryu:2006bv}
  S.~Ryu and T.~Takayanagi,
  Phys.\ Rev.\ Lett.\  {\bf 96} (2006) 181602
  [hep-th/0603001].

\bibitem{Stanford:2014jda}
  D.~Stanford and L.~Susskind,
  Phys.\ Rev.\ D {\bf 90} (2014) no.12,  126007
  [arXiv:1406.2678 [hep-th]].

\bibitem{Susskind:2014rva}
  L.~Susskind,
  Fortsch.\ Phys.\  {\bf 64}, 24 (2016)
  [arXiv:1403.5695 [hep-th], arXiv:1402.5674 [hep-th]].


\bibitem{Brown:2015lvg}
  A.~R.~Brown, D.~A.~Roberts, L.~Susskind, B.~Swingle and Y.~Zhao,
  Phys.\ Rev.\ D {\bf 93}, no. 8, 086006 (2016)
  [arXiv:1512.04993 [hep-th]].

\bibitem{Brown:2015bva}
  A.~R.~Brown, D.~A.~Roberts, L.~Susskind, B.~Swingle and Y.~Zhao,
  Phys.\ Rev.\ Lett.\  {\bf 116}, no. 19, 191301 (2016)
  [arXiv:1509.07876 [hep-th]].

\bibitem{Cai:2016xho}
  R.~G.~Cai, S.~M.~Ruan, S.~J.~Wang, R.~Q.~Yang and R.~H.~Peng,
  JHEP {\bf 1609}, 161 (2016)
  [arXiv:1606.08307 [gr-qc]].

\bibitem{Huang:2016fks}
  H.~Huang, X.~H.~Feng and H.~Lu,
  Phys.\ Lett.\ B {\bf 769}, 357 (2017)
  [arXiv:1611.02321 [hep-th]].

\bibitem{Pan:2016ecg}
  W.~J.~Pan and Y.~C.~Huang,
  Phys.\ Rev.\ D {\bf 95}, no. 12, 126013 (2017)
  [arXiv:1612.03627 [hep-th]].

\bibitem{Guo:2017rul}
  W.~D.~Guo, S.~W.~Wei, Y.~Y.~Li and Y.~X.~Liu,
  Eur.\ Phys.\ J.\ C {\bf 77}, no. 12, 904 (2017)
  [arXiv:1703.10468 [gr-qc]].

\bibitem{Abad:2017cgl}
  F.~J.~G.~Abad, M.~Kulaxizi and A.~Parnachev,
  JHEP {\bf 1801}, 127 (2018)
  [arXiv:1705.08424 [hep-th]].

\bibitem{Lehner:2016vdi}
  L.~Lehner, R.~C.~Myers, E.~Poisson and R.~D.~Sorkin,
  Phys.\ Rev.\ D {\bf 94} (2016) no.8,  084046
  [arXiv:1609.00207 [hep-th]].

\bibitem{Jubb:2016qzt}
  I.~Jubb, J.~Samuel, R.~Sorkin and S.~Surya,
  Class.\ Quant.\ Grav.\  {\bf 34}, no. 6, 065006 (2017)
  [arXiv:1612.00149 [gr-qc]].

\bibitem{Carmi:2016wjl}
  D.~Carmi, R.~C.~Myers and P.~Rath,
  JHEP {\bf 1703}, 118 (2017)
  [arXiv:1612.00433 [hep-th]].

\bibitem{Reynolds:2016rvl}
  A.~Reynolds and S.~F.~Ross,
  Class.\ Quant.\ Grav.\  {\bf 34}, no. 10, 105004 (2017)
  [arXiv:1612.05439 [hep-th]].

\bibitem{Cai:2017sjv}
  R.~G.~Cai, M.~Sasaki and S.~J.~Wang,
  Phys.\ Rev.\ D {\bf 95}, no. 12, 124002 (2017)
  [arXiv:1702.06766 [gr-qc]].

\bibitem{Tao:2017fsy}
  J.~Tao, P.~Wang and H.~Yang,
  Eur.\ Phys.\ J.\ C {\bf 77}, no. 12, 817 (2017)
  [arXiv:1703.06297 [hep-th]].

\bibitem{Wang:2017uiw}
  P.~Wang, H.~Yang and S.~Ying,
  Phys.\ Rev.\ D {\bf 96}, no. 4, 046007 (2017)
  [arXiv:1703.10006 [hep-th]].

\bibitem{Miao:2017quj}
  Y.~G.~Miao and L.~Zhao,
  Phys.\ Rev.\ D {\bf 97}, no. 2, 024035 (2018)
  [arXiv:1708.01779 [hep-th]].

\bibitem{Jiang:2018sqj}
  J.~Jiang and H.~Zhang,
  arXiv:1806.10312 [hep-th].

\bibitem{Jiang:2018pfk}
  J.~Jiang,
  Phys.\ Rev.\ D {\bf 98}, no. 8, 086018 (2018)
  [arXiv:1810.00758 [hep-th]].

\bibitem{Meng:2018vtl}
  K.~Meng,
  arXiv:1810.02208 [hep-th].

\bibitem{Brown:2018bms}
  A.~R.~Brown, H.~Gharibyan, H.~W.~Lin, L.~Susskind, L.~Thorlacius and Y.~Zhao,
  arXiv:1810.08741 [hep-th].

\bibitem{Jackiw:2003pm}
  R.~Jackiw and S.~Y.~Pi,
  Phys.\ Rev.\ D {\bf 68}, 104012 (2003)
  [gr-qc/0308071].

\bibitem{Alexander:2009tp}
  S.~Alexander and N.~Yunes,
  Phys.\ Rept.\  {\bf 480}, 1 (2009)
  [arXiv:0907.2562 [hep-th]].

\bibitem{Grumiller:2008ie}
  D.~Grumiller, R.~B.~Mann and R.~McNees,
  Phys.\ Rev.\ D {\bf 78}, 081502 (2008)
  [arXiv:0803.1485 [gr-qc]].

\bibitem{Saremi:2011ab}
  O.~Saremi and D.~T.~Son,
  JHEP {\bf 1204}, 091 (2012)
  [arXiv:1103.4851 [hep-th]].

\bibitem{Fischler:2015kro}
  W.~Fischler and S.~Kundu,
  JHEP {\bf 1604}, 112 (2016)
  [arXiv:1512.01238 [hep-th]].

\bibitem{Zhao:2015inu}
  T.~Y.~Zhao and T.~Wang,
  JCAP {\bf 1606}, no. 06, 019 (2016)
   [arXiv:1512.01919 [gr-qc]].

\bibitem{Abbott:2018utx}
  B.~P.~Abbott {\it et al.} [LIGO Scientific and Virgo Collaborations],
  Phys.\ Rev.\ Lett.\  {\bf 120}, no. 20, 201102 (2018)
  [arXiv:1802.10194 [gr-qc]].

\bibitem{Loutrel:2018ydv}
  N.~Loutrel, T.~Tanaka and N.~Yunes,
  Phys.\ Rev.\ D {\bf 98}, no. 6, 064020 (2018)
  [arXiv:1806.07431 [gr-qc]].

\bibitem{Gao:2018acg}
  Y.~X.~Gao, Y.~Huang and D.~J.~Liu,
  arXiv:1808.01433 [gr-qc].

\bibitem{Tahura:2018zuq}
  S.~Tahura and K.~Yagi,
  Phys.\ Rev.\ D {\bf 98}, no. 8, 084042 (2018)
  [arXiv:1809.00259 [gr-qc]].

\bibitem{Nishizawa:2018srh}
  A.~Nishizawa and T.~Kobayashi,
  arXiv:1809.00815 [gr-qc].

\bibitem{Yagi:2012ya}
  K.~Yagi, N.~Yunes and T.~Tanaka,
  Phys.\ Rev.\ D {\bf 86}, 044037 (2012)
  Erratum: [Phys.\ Rev.\ D {\bf 89}, 049902 (2014)]
  [arXiv:1206.6130 [gr-qc]].

\bibitem{Alexander:2004xd}
  S.~H.~S.~Alexander and S.~J.~Gates, Jr.,
  JCAP {\bf 0606} (2006) 018
  [hep-th/0409014].

\bibitem{Campbell:1990fu}
  B.~A.~Campbell, M.~J.~Duncan, N.~Kaloper and K.~A.~Olive,
  Nucl.\ Phys.\ B {\bf 351}, 778 (1991).

\end{thebibliography}
\end{document}